\documentclass[footinbib,aps,prapplied,reprint,superscriptaddress,nopacs,longbibliography]{revtex4-2}

\usepackage{amsmath,amssymb,bm}
\usepackage{physics,graphicx,tabularx,braket}
\usepackage{xcolor}
\usepackage{array}
\usepackage{multirow}
\usepackage[absolute]{textpos}

\definecolor{darkblue}{rgb}{0,0,0.5}
\usepackage{hyperref}
\hypersetup{
  colorlinks=true,
  linkcolor=black,
  filecolor=blue,
  citecolor=darkblue,
  urlcolor=black,
}

\definecolor{joeBlue}{RGB}{0,163,224}

\newcolumntype{Y}{>{\centering\arraybackslash}p{1.35cm}} 
\newcolumntype{Z}{>{\centering\arraybackslash}p{1.75cm}} 

\begin{document}

\title{Photonic Links for Spin-Based Quantum Sensors}

\author{M. Reefaz Rahman}
\affiliation{Department of Electrical and Computer Engineering, The University of Alabama, Tuscaloosa, Alabama 35487, USA}

\author{Karsten Schnier}
\affiliation{Department of Electrical and Computer Engineering, The University of Alabama, Tuscaloosa, Alabama 35487, USA}

\author{Ryan Goldsmith}
\affiliation{Department of Electrical and Computer Engineering, The University of Alabama, Tuscaloosa, Alabama 35487, USA}

\author{Benjamin J. Lawrie}
\affiliation{Materials Science and Technology Division, Oak Ridge National Laboratory, Oak Ridge, Tennessee 37831, USA}

\author{Joseph M. Lukens}
\affiliation{Elmore Family School of Electrical and Computer Engineering and Purdue Quantum Science and Engineering Institute, Purdue University, West Lafayette, Indiana 47907, USA}
\affiliation{Quantum Information Science Section, Oak Ridge National Laboratory, Oak Ridge, Tennessee 37831, USA}

\author{Seongsin M. Kim}
\affiliation{Department of Electrical and Computer Engineering, The University of Alabama, Tuscaloosa, Alabama 35487, USA}

\author{Patrick Kung}
\email{patkung@eng.ua.edu}
\affiliation{Department of Electrical and Computer Engineering, The University of Alabama, Tuscaloosa, Alabama 35487, USA}

\begin{abstract}

A growing variety of optically accessible spin qubits have emerged in recent years as key components for quantum sensors, qubits, and quantum memories. However, the scalability of conventional spin-based quantum architectures remains limited by direct microwave delivery, which introduces thermal noise, electromagnetic cross-talk, and design constraints for cryogenic, high-field, and distributed systems. In this work, we present a unified framework for RF-over-fiber (RFoF) control of optically accessible spins through RFoF optically detected magnetic resonance (ODMR) spectroscopy of nitrogen-vacancy (NV) centers in diamond. The RFoF platform relies on an electro-optically modulated telecom-band laser that transmits microwave signals over fiber and a high-speed photodiode that recovers the RF waveform to drive NV center spin transitions. We obtain an RFoF efficiency of 1.81\% at 2.90~GHz, corresponding to $P_{\mathrm{RF,out}}=-0.7$~dBm. The RFoF architecture provides a path toward low-noise, thermally isolated, and cryo-compatible ODMR systems bridging conventional spin-based quantum sensing protocols with emerging distributed quantum technologies.

\end{abstract}

\maketitle


\section{Introduction}

Optically accessible spin defects like the nitrogen-vacancy (NV) center in diamond are widely used quantum sensors because their spin state can be initialized and read out optically across a wide range of temperatures, enabling measurements of magnetic fields, strain/electric fields, and temperature across nanometer to centimeter length scales \cite{doherty,Atature2018Review,Quantumguidelines,wu2025nanoscale}. Over the past decade, other spin defects like the boron vacancy in hBN \cite{JacquesthinhBN,ChunhuithinhBN,Gottscholl2021SpinDefects,solanki2025sub,vlassiouk2025defect} and group IV defects in diamond \cite{GroupIVdefects,hepp2014electronic,wu2025coherent,wu2025experimental} have drawn increasing interest for applications in quantum sensing and quantum networking, respectively. Multi-node NV-center quantum networks have demonstrated entanglement swapping across three nodes \cite{pompili_2021_science_multinode,hermans_2022_nature_teleport}, and metropolitan-scale heralded entanglement between independently operated nodes separated by $10~\mathrm{km}$ has been reported using deployed optical fiber and phase-stabilized architectures \cite{stolk_2024_sciadv_10km}. These results motivate experimental architectures in which control, frequency references, and timing can be distributed robustly over distance while remaining compatible with high magnetic fields and cryogenic environments.

However, despite substantial progress in the control of optically accessible spins across a growing variety of material platforms, the scalability of the rf delivery needed to coherently drive spin transitions remains a substantial thermal-engineering challenge in cryogenic and high-field environments. This challenge has been highlighted by nascent sub-Kelvin scanning NV microscopy experiments that are often constrained by the thermal load associated with rf power delivery \cite{scheidegger_2022_apl_350mk}. A more substantial challenge arises in high-field cryogenic experiments: as the magnetic field parallel to the spin axis increases, the relevant spin transitions for the NV center quickly scale to frequencies that are impossible to access with conventional cryogenic coaxial cables. While there has been some initial success in room-temperature NV-detected NMR experiments at fields of $4.2~\mathrm{T}$ corresponding to spin transition frequencies of $115~\mathrm{GHz}$ \cite{ren2023demonstration,fortman_2020_apl_115ghz}, scaling efforts at these frequencies to cryogenic temperatures introduces enormous technical challenges. These challenges are highlighted by recent cryogenic 7 T probes of hBN boron vacancy spin defects\cite{solanki2025sub} that relied on all optical spin control due to the challenges associated with sub-THz microwave delivery into cryogenic environments. The ability to coherently address spins in these environments would unlock substantial new opportunities in spin based quantum sensing.

To address the microwave-delivery bottleneck for high-frequency, high-field, and cryogenic spin-based quantum systems and to enable link-compatible architectures, we adopt an RF-over-fiber (RFoF) approach. In RFoF links, microwave signals are carried on an optical carrier, transported through optical fiber, and recovered electrically using a high-speed photodiode placed near the experiment. RFoF links have notably been used for coherent control and readout of superconducting qubits \cite{lecocq2021control}, but substantially larger rf powers are required to drive spin transitions in spin defects like the NV center. Two RFoF approaches are especially relevant here: (i) optical heterodyne generation, where two lasers with a controlled frequency offset beat on a photodiode to generate a tunable microwave or sub-THz carrier, and (ii) electro-optic modulator (EOM)-based links, where a stable RF source drives an EOM and the modulated optical carrier is delivered over fiber and converted back to RF at the remote photodiode \cite{xu_2014_prj_mwp,rinaldi_2025_aplphotonics_rof}. RFoF and microwave-photonics frequency-transfer architectures have demonstrated phase-stable distribution of millimeter-wave carriers near $100~\mathrm{GHz}$ over long fiber links \cite{deng_2018_oe_100ghz,deng_2020_oe_multifreq,chen_2025_nsr_timefreq}, and modern photonic microwave-generation work highlights heterodyne and photomixing approaches as standard routes to low-noise high-frequency carriers \cite{sun_2024_nature_ofd}. RFoF has also been developed for high magnetic-field MRI receive systems, where dynamic range and safety constraints are stringent \cite{fan_2021_rsi_rfof_mri,yuan_2007_jmr_mriopt}.

In this work, we demonstrate proof-of-principle EOM-based RFoF control of NV center optically detected magnetic resonance (ODMR) at room temperature in low field environments, and we benchmark these measurements against conventional rf delivery as a function of rf power. For the RFoF implementation, the recovered microwave tone delivered to the antenna reaches $P_{\mathrm{RF,out}}=-0.7$~dBm near 2.90~GHz; with an efficiency of 1.81\%. At this RFoF operating point we observe $>2.2$\% ODMR contrast at a fixed magnet position, while for a field sweep of 8--36~G we obtain 0.6--0.9\% contrast at $P_{\mathrm{RF,out}}=-5.5$~dBm. With this proof of principle in hand, we explore how RFoF delivery provides a scalable pathway toward future high-field (multi-10~GHz to sub-THz), cryogenic, and networked-node operation \cite{ren2023demonstration,fortman_2020_apl_115ghz,deng_2018_oe_100ghz,deng_2020_oe_multifreq,chen_2025_nsr_timefreq,pompili_2021_science_multinode,hermans_2022_nature_teleport,stolk_2024_sciadv_10km,scheidegger_2022_apl_350mk,li_2024_oe_optical_mw_ctrl}.


\section{Background and Experimental Details}

\begin{figure}[hb]
    \centering
    \includegraphics[width=\columnwidth]{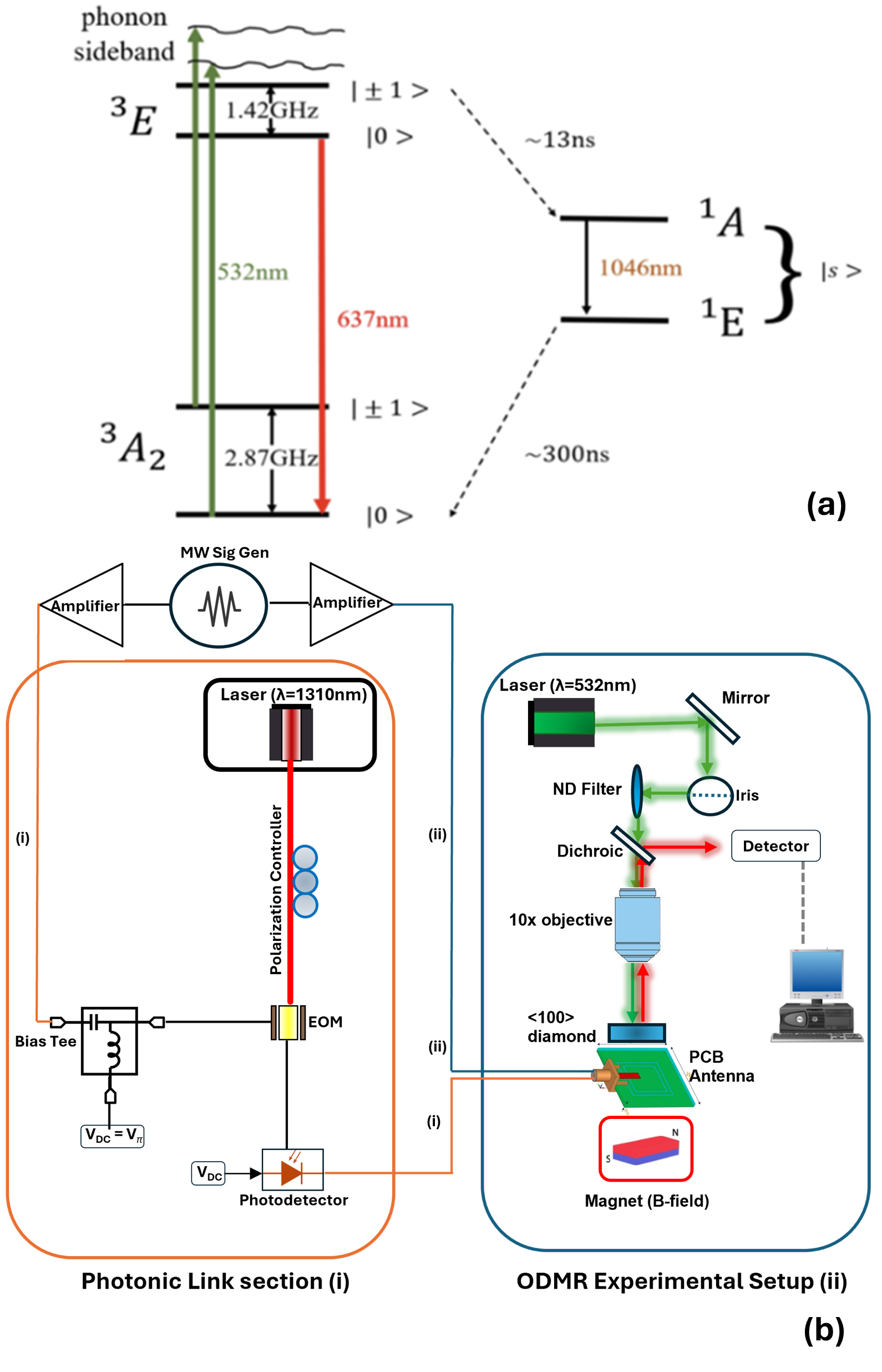}
\caption{Overview of NV-center ODMR physics and the experimental architectures used in this work. (a) Simplified energy-level diagram of the negatively charged NV center, illustrating optical spin polarization under 532~nm excitation, spin-dependent fluorescence near 637~nm, and shelving through intermediate singlet states that enables ODMR contrast; the ground-state spin transition is split by $D\approx 2.87$~GHz at zero field. (b) Experimental architectures used in this work: (i) RFoF microwave delivery chain in which a 1310~nm laser is electro-optically modulated by an RF source (with optional amplification), transmitted over fiber, and converted back to an electrical microwave tone by a high-speed photodiode biased via a bias-tee/SMU before driving the antenna near the NV sample; (ii) conventional free-space ODMR configuration used as a baseline, including optical excitation/collection optics, a microwave antenna adjacent to the diamond, and an external magnet to apply a controllable field.}
\label{fig:overview}
\end{figure}

\subsection{Theoretical Model and Geometric Considerations for ODMR Measurements}

The negatively charged nitrogen-vacancy (NV$^-$) center in diamond is a point defect with a spin-triplet ground state ($^3A_2$) and an optically accessible spin-triplet excited state ($^3E$), with intermediate singlet states that enable spin-dependent non-radiative decay, as illustrated in Fig.~\ref{fig:overview}(a). Under 532~nm excitation, repeated optical cycling polarizes the ground state into $m_s=0$. Resonant microwave driving transfers the population to $m_s=\pm1$, increasing the probability of intersystem crossing and reducing photoluminescence (PL), producing an ODMR spectrum as the RF frequency is swept across the spin transitions.

To first order, the ground-state spin Hamiltonian can be written as
\begin{equation}
    H = D S_z^2 + E(S_x^2 - S_y^2) + \gamma_e \mathbf{B}\cdot\mathbf{S},
\end{equation}
where $D \approx 2.87~\mathrm{GHz}$ is the zero-field splitting, $E$ accounts for the strain-induced splitting, $\gamma_e \approx 2.8~\mathrm{MHz/G}$ is the electron gyromagnetic ratio, $\mathbf{B}$ is the applied magnetic field, and $\mathbf{S}$ is the spin-1 operator. For a given NV axis, the ODMR transition frequencies are approximately
\begin{equation}
    f_{\pm}(B,\alpha) \approx D \pm \gamma_e B \cos\alpha,
\end{equation}
where $\alpha$ is the angle between the applied field and the NV axis.

For all ODMR spectra, the contrast is computed as
\begin{equation}
    C = \frac{I_{\mathrm{off}} - I_{\mathrm{on}}}{I_{\mathrm{off}}},
\end{equation}
where $I_{\mathrm{off}}$ and $I_{\mathrm{on}}$ are the integrated PL signals for microwave-off and microwave-on conditions, respectively, using the same optical excitation and collection settings. Resonance linewidths are reported as the full width at half maximum (FWHM) extracted from fitted ODMR spectra.

The dependence of contrast and linewidth on RF power follows standard driven-spin behavior: increasing the RF drive increases the transition rate (and thus contrast) until saturation, while strong driving can broaden the resonance (power broadening) and modify the apparent line shape. These effects motivate a direct, power-referenced comparison between conventional coaxial delivery and RFoF delivery.

\subsection{ODMR Setup and Measurement Procedure}

All measurements were performed on a single-crystal electronic-grade CVD diamond (Thorlabs DNVB1; $5 \times 5~\mathrm{mm}^2$, 300~\textmu m thickness), cut along the $\langle100\rangle$ axis with an NV$^-$ concentration of approximately 300~ppb. Optical excitation was provided by a continuous-wave 532~nm laser. The beam was attenuated using neutral-density filters and focused through a 10$\times$ microscope objective (NA~$\approx$~0.25) to a $\sim$$20~\mu$m spot on the diamond surface. Fluorescence (550--850~nm) was collected through the same objective and imaged onto a CCD camera. For contrast calculations, the PL was integrated consistently across microwave-on/off acquisitions. 

For baseline ODMR measurements, the microwave excitation was generated by a signal generator. ODMR spectra were acquired by sweeping the frequency across the NV resonance while recording the PL intensity for microwave-on and microwave-off conditions, with averaging over multiple cycles to improve the signal-to-noise ratio. Resonance frequencies, linewidths, and contrasts were extracted by fitting the measured spectra. 

\subsection{RF Delivery Paths and Power Reference Planes}

Two microwave delivery paths illustrated in Fig.~\ref{fig:overview}(b) were evaluated. In the \textit{conventional} configuration, the RF signal was delivered via a coaxial cable directly to the antenna; the applied RF power reported for this baseline corresponds to the power delivered to an antenna feedpoint (up to 25~dBm in this work). In the RFoF configuration, the recovered RF tone from the photodiode output was routed to the same antenna; the RF power reported for RFoF corresponds to the recovered RF power delivered to the antenna feedpoint (up to $P_{\mathrm{RF,out}}=-0.7$~dBm near 2.90~GHz). This explicit reference-plane definition enables a direct comparison of ODMR contrast and linewidth as a function of delivered RF power for the two delivery methods. However, in future cryogenic, high-field implementations, the RFoF implementation offers favorable power scaling because the relevant RFoF power is measured at the output of the cryogenic photodetector inside the cryostat, thus avoiding the losses associated with conventional coaxial delivery into cryogenic environments. For RFoF link characterization, we also report an optical-to-RF power conversion efficiency based on the optical power incident on the photodiode ($P_{\mathrm{opt,PD}}$) and the recovered RF power delivered to the antenna.

\subsection{RF-over-Fiber Microwave Delivery}
\label{subsec:rfof_methods}

In the RFoF configuration, a continuous-wave 1310~nm laser served as the optical carrier. The laser polarization was adjusted with a fiber polarization controller before entering a Mach--Zehnder electro-optic modulator (EOM). The RF signal from the signal generator was combined with a DC bias from a source meter using a bias tee and applied to the EOM to intensity-modulate the optical carrier. The modulated light propagated through single-mode fiber and the RF tone was recovered by a high-speed photodiode. The recovered RF signal then drove the same PCB antenna used in the direct-feed coax baseline configuration, enabling direct comparison between coaxial delivery and RFoF delivery under matched optical detection conditions.

\subsection{Microwave Antenna Design and Characterization}

A broadband microstrip "pinhole'' antenna was employed to generate the microwave magnetic field required for ODMR. The antenna layout follows the broadband, large-area design reported by Sasaki \textit{et al.} \cite{sasaki2016rsi_antenna}, which concentrates the microwave magnetic field near the aperture region while maintaining broadband operation around the NV zero-field splitting (2.87~GHz). In our implementation, the antenna was fabricated on an FR-4 substrate (1.6~mm thickness) with a copper top layer and metallic ground plane, and the diamond was positioned above the aperture to maximize coupling to the NV spin transitions. The antenna response was verified by measuring the reflection coefficient ($S_{11}$), confirming broadband coupling in the 2.8--3.0~GHz range. The same antenna and sample placement were used for both the coaxial ODMR baseline and RFoF measurements to enable a direct comparison of ODMR response versus delivered RF power.


\begin{figure}[hbt!]
    \centering
    \includegraphics[width=0.95\columnwidth]{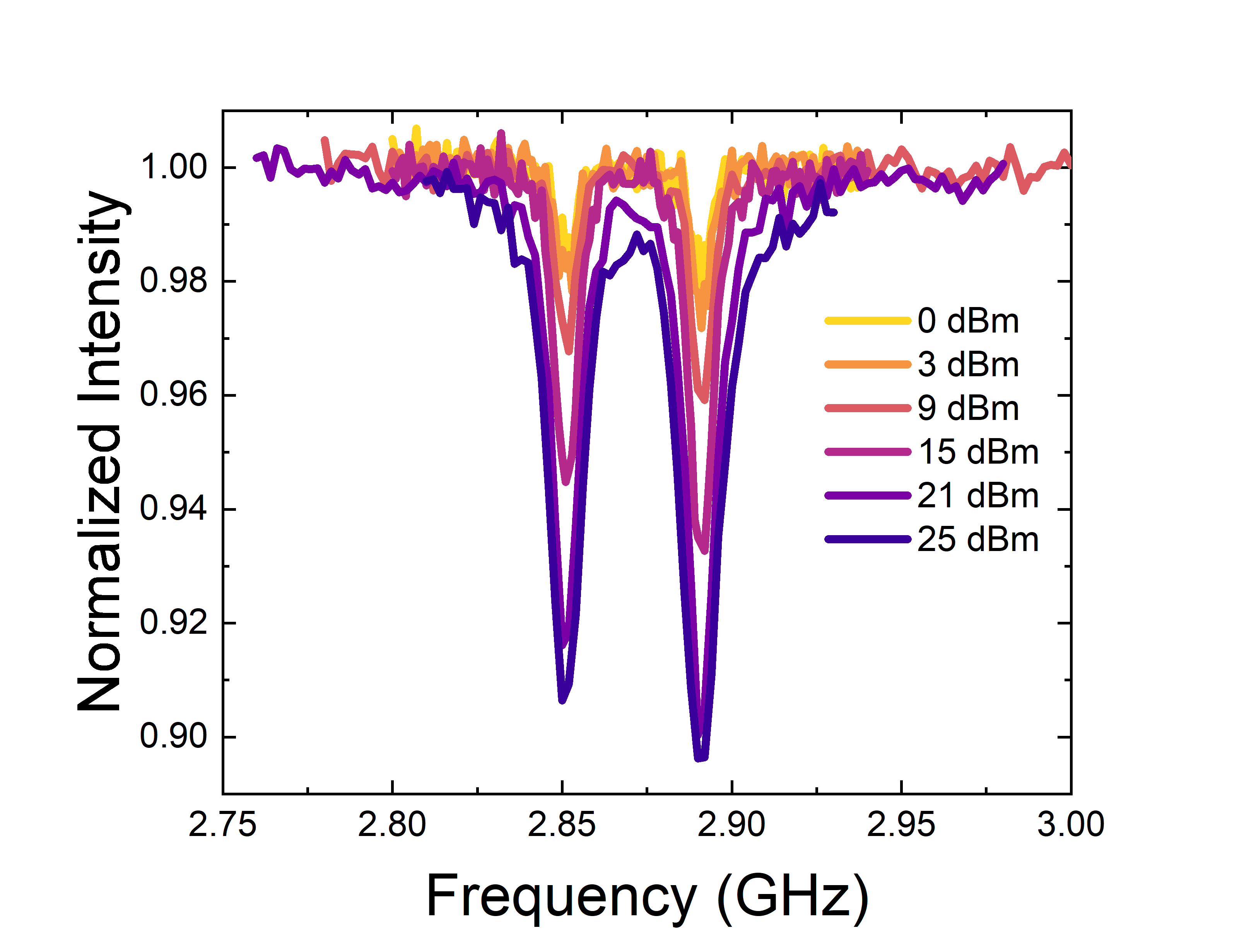}
    \caption{Power dependent ODMR spectra for ensemble NV centers in diamond using the conventional coax-fed antenna. ODMR spectra were measured at increasing RF power levels (0, 3, 9, 15, 21, and 25~dBm) delivered to the antenna feedpoint, with normalized photoluminescence plotted versus microwave frequency. Increasing RF power increases the ODMR contrast and produces modest linewidth broadening consistent with stronger driving. Optical settings were held constant during the sweep.}
    \label{fig:RFpower}
\end{figure}

\section{Results and Discussion}
\label{sec:results}

\subsection{Baseline Power-Dependent ODMR Spectrum}

Baseline conventional ODMR spectra were recorded while varying the delivered RF power from 0~dBm to 25~dBm under identical optical conditions. The diamond sample and antenna assembly were held fixed in a 11.2~G field. Unless otherwise noted, the reported RF powers in this subsection refer to the power delivered to the antenna feedpoint in the conventional coax-fed configuration.

Figure~\ref{fig:RFpower} shows normalized ODMR spectra acquired at increasing RF power. The resonances near 2.87~GHz correspond to the $\ket{m_s = 0} \leftrightarrow \ket{m_s = \pm1}$ transitions of the NV ground state (with Zeeman splitting set by the applied field and its projection onto the NV axes). Two consistent trends are observed as RF power increases: (i) the contrast increases, and (ii) the linewidths broaden modestly. At 0~dBm, the contrast is approximately 2\%, while at 25~dBm it exceeds 11\%. The increase in contrast reflects stronger spin driving and enhanced population transfer, whereas the linewidth increase is consistent with power broadening under strong driving.

This contrast--linewidth trade-off aligns with prior NV-ODMR studies, where increasing Rabi frequency initially improves ODMR visibility but excessive drive reduces spectral resolution due to power broadening and additional dephasing mechanisms \cite{Taylor2008,Maze2008,Pham2011,Barry2020,Dreau2011}. In our conventional configuration, 25~dBm provides the highest contrast and is used as the baseline reference point for comparison to RF-over-fiber measurements in Sec.~\ref{subsec:rfof_results}.

\subsection{RFoF ODMR Measurement and Results}
\label{subsec:rfof_results}

To evaluate RFoF delivery for ODMR interrogation, we modulate an optical carrier at 1310~nm using an EOM (see Sec.~\ref{subsec:rfof_methods} in Experimental Details). The recovered RF signal at the photodiode is delivered to the same broadband antenna used in the coax baseline, enabling a direct comparison based on delivered RF power and identical optical detection settings.

\begin{figure}[!htb]
    \centering
    \includegraphics[width=0.95\columnwidth]{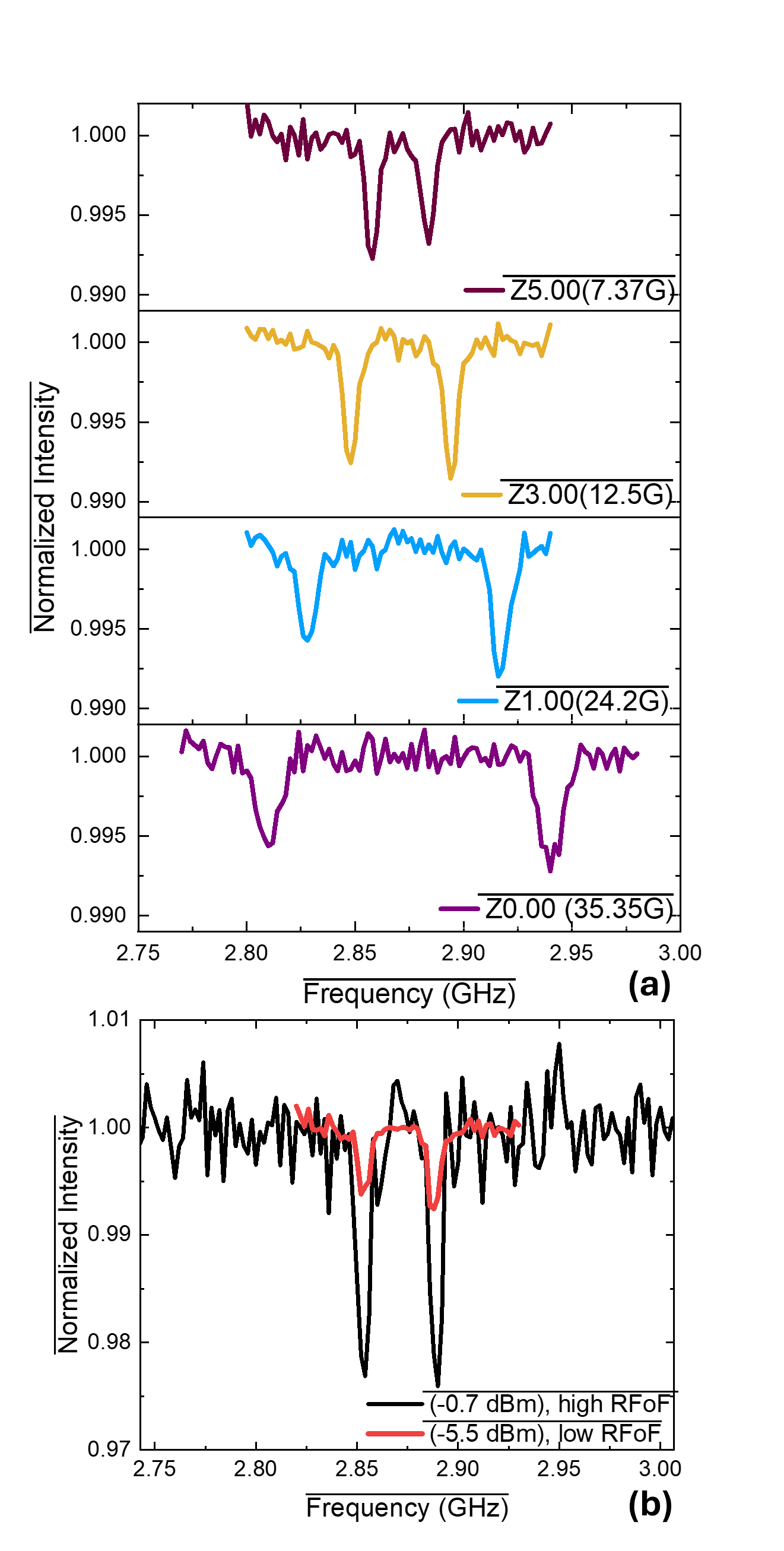}
    \caption{RFoF ODMR spectra obtained using the recovered microwave tone from the photodiode to drive the NV ensemble through the same broadband antenna used in the coaxial baseline. (a) Stacked RFoF-driven ODMR spectra near the zero-field splitting ($\sim$2.87~GHz) for varying magnetic fields of 8--36~G, measured at $P_{\mathrm{RF,out}}=-5.5$~dBm delivered to the antenna; the resonance splitting increases with field in agreement with the Zeeman effect and the observed contrast is 0.6--0.9\%. (b) RFoF-driven ODMR at a fixed magnetic field comparing two recovered RF powers, showing enhanced contrast at higher delivered power; at $P_{\mathrm{RF,out}}=-0.7$~dBm (2.90~GHz) the contrast exceeds 2.2\%, while the lower-power condition reproduces the characteristic ODMR spectrum near 2.87~GHz.}
    \label{fig:rfof_odmr}
\end{figure}

Figure~\ref{fig:rfof_odmr}(a) shows RFoF-driven ODMR spectra acquired across magnetic fields spanning 8--36~G. The ODMR splitting increases monotonically with magnetic-field strength, consistent with Zeeman splitting. At $P_{\mathrm{RF,out}}=-5.5$~dBm delivered to the antenna, the measured RFoF ODMR contrast ranges between 0.6\% and 0.9\% across the field sweep. Figure~\ref{fig:rfof_odmr}(b) highlights the RF-power dependence of RFoF-driven ODMR at a fixed magnetic field. Increasing the recovered RFoF drive to $P_{\mathrm{RF,out}}=-0.7$~dBm near 2.90~GHz increases the ODMR contrast to $>2.2$\%. 

Even with substantially lower delivered RF power than the 25~dBm coax-fed baseline, the RFoF link reliably reproduces the characteristic ODMR response and enables controlled spin driving through fiber-delivered microwaves.
We additionally quantify the RFoF link performance using an optical-to-RF power conversion efficiency defined as
\begin{equation}
\eta_{O\rightarrow RF}=\frac{P_{\mathrm{RF,ant}}}{P_{\mathrm{opt,PD}}},
\end{equation}
where $P_{\mathrm{opt,PD}}$ is the optical power incident on the photodiode and $P_{\mathrm{RF,ant}}$ is the recovered RF power delivered to the antenna feedpoint. For the highest-power RFoF operating point in Fig.~\ref{fig:rfof_odmr}(b), we measure $P_{\mathrm{opt,PD}}=47~\mathrm{mW}$ and $P_{\mathrm{RF,ant}}=-0.7$~dBm at 2.90~GHz, corresponding to $\eta_{O\rightarrow RF}=0.0181$ (1.81\%).

The reduced contrast at lower RFoF delivered power is primarily attributed to insertion losses and finite modulation depth in the optical link, which limit recovered RF amplitude at the antenna relative to direct coaxial delivery. Nonetheless, RFoF provides electrical isolation between the RF source and the sensing node, reducing electromagnetic interference and enabling architectures compatible with thermally isolated and cryogenic environments. Further improvements in modulation depth, photodiode linearity, and optical power handling are expected to increase recovered RF power and improve contrast while preserving the advantages of fiber distribution.

Overall, these measurements establish RF power as a primary control parameter for ODMR contrast and linewidth in both delivery architectures, and show that an RFoF link can reliably reproduce the ODMR spectrum while providing a pathway to electrically isolated and thermally compatible microwave delivery for future high-field and cryogenic operation.

\section{Conclusion}

We have presented a concise proof-of-principle demonstration of RFoF control of NV centers in diamond using RFoF microwave delivery at 1310~nm with EOM-based modulation and photodiode recovery. The RFoF link reproduces the characteristic ODMR spectrum near 2.87~GHz, yielding 0.6--0.9\% contrast for an 8--36~G field sweep at $P_{\mathrm{RF,ant}}=-5.5$~dBm, and exceeding 2.2\% contrast at a fixed field for a higher delivered power of $P_{\mathrm{RF,ant}}=-0.7$~dBm. For this operating point, we measure an optical-to-RF power conversion efficiency of $\eta_{O\rightarrow RF}=0.0181$ (1.81\%), defined at the antenna feedpoint with the long SMA-cable loss included.

These results establish RFoF as a practical microwave distribution approach for NV-ODMR that preserves the ODMR spectral signature while providing electrical isolation between the source and sensing node. Future work will focus on increasing recovered RF power and link linearity through improved modulation depth, higher optical power handling, and optimized photodiode operation. The approach used here is scalable to higher frequency operation of order 100 GHz\cite{multani2024quantum}, enabling the future RFoF coherent control of optically accessible spins in cryogenic high-field environments that have otherwise been impossible to access with conventional techniques. 

\section{Acknowledgments}
This material is based upon work supported by the U.S. Department of Energy, Office of Science, Office of Basic Energy Sciences under Award Number DE-SC0025432. Research at ORNL was sponsored by the U. S. Department of Energy, Office of Science, Basic Energy Sciences, Materials Sciences and Engineering Division.

\bibliography{Refs}

\end{document}